\documentclass[twocolumn,twoside,slac_two]{revtex4}
\usepackage{graphicx}
\usepackage{fancyhdr}
\def\fermilat{{\em Fermi}/LAT}
\def\fermi{{\em Fermi}}
\def\twocms{2{\,}cm{~}Survey}
\def\moj{MOJAVE}

\begin{document}

\title{Radio and Gamma Properties of the 2\,cm Survey and MOJAVE Samples}

%

\author{
Eduardo Ros for the MOJAVE collaboration
}
\affiliation{
Departament d'Astronomia i Astrof\'{\i}sica, Universitat de Val\`encia, E-46100 Burjassot, Spain\\
Max-Planck-Institut f\"ur Radioastronomie, Auf dem H\"ugel 69, D-53121 Bonn, Germany
}

\begin{abstract}
The 2\,cm~VLBA Survey observed since 1994 a set of $\sim$170 Quasars, BL Lac objects, and radio galaxies, 
selected to be representative of the compact AGN radio population.  This effort was continued as the 
MOJAVE project, where a statistically complete set of radio sources being monitored was defined.  A comparison of the gamma-detection rates between
the members of both samples shows that 
the MOJAVE-I sources, hosting generally faster jets, have a 
much higher detection rate than the sources not belonging to this sample.  
BL\,Lac objects are more favourably detected than QSOs in
gamma-rays, in the same rate for both samples.

\end{abstract}

\maketitle

\thispagestyle{fancy}


\section{Background}
The \twocms\ observed from 1994 to 2002 a set of 
171 Quasars, BL Lac objects, and radio galaxies, selected to be 
representative of the compact AGN radio population 
(see e.g., \cite{kel04}; images in \cite{kel98} and \cite{zen02};
kinematical results of the parsec-scale jet features in \cite{kel04}
and \cite{ros10}).
This sample was redefined to include 135 objects 
(two thirds from those belonging to the \twocms\ sample) and to be
statistically complete  with continued monitoring observations 
from 2002 with the name of the \moj-1 program 
(see \cite{lis09a} 
for a description of the images, 
and \cite{lis09c,hom09}
for kinematical results). 

The Large Area Telescope (LAT, see \cite{atw09}) on board
the {\em Fermi Gamma-ray Space Telescope} has detected a big fraction
of the AGN present in these samples.  
Furthermore, \cite{kov09b} has shown that the majority of the \fermi\ blazars 
are radio-loud and show a core-jet structure at parsec scales 

First results on the gamma-radio relationship were published for
the statistically complete sample, in a first instance for the LAT
Bright AGN Source (LBAS) 3-month list (see \cite{kov09,lis09b,pus09,sav09}
for the \moj\ results, and \cite{abd09a} for the LBAS).  The
results presented in this conference are based on a preliminary
11-month list of \fermilat\ detections as of early November 2009, at the
time of the Fermi Symposium 2009.  

From the \twocms\ sample, 79 prominent AGN are not members of the 
statistically complete sample (\moj-I), and their radio properties
can be as well compared to the \fermilat\ findings (19 of those are detected
at the preliminary 11-month list).

\section{Observations}

Kinematic values of the AGN jets were obtained from the intensive
monitoring survey being performed since 1994.
Generally, the Very Long Baseline Array (VLBA) observed
each AGN for a total time of 50--60\,min in scans of several
minutes spread over 8\,hr, providing
an almost full interferometric track.  Each source was observed over
several epochs during those years, with a time sampling of at least
one observation per year.
Additional, high-quality VLBA archival epochs were
added to the data base.

After (semi-automatic) imaging of the VLBI observations,
the features observed in the jets were modeled by Gaussian 
functions fitted to the interferometric visibilities.  
Those features were identified over several epochs, and the measurement
of their relative positions provided kinematical values from which
further statistical studies can be performed, as well as individual
source studies (see \cite{lis09b} for a description of the
imaging process and \cite{lis09c,hom09} and references therein
for kinematic results of the \moj-I sample, and \cite{kel04,ros10} for
the kinematic analysis of the \twocms\ sample.).
Here we compare these results with the \fermilat\ detections.


\begin{figure}[b!]
\includegraphics[width=65mm]{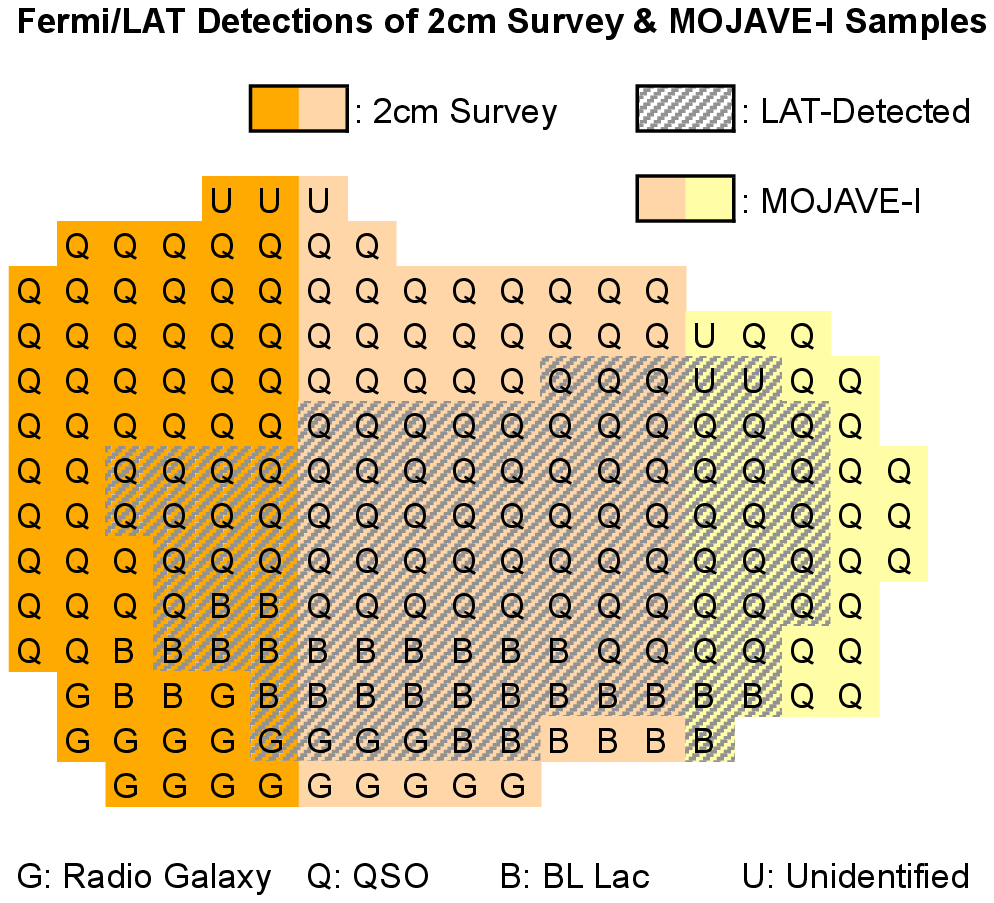}
\caption{\fermilat\ Detections of \twocms\ and \moj\
Sources, as of November 2009 (preliminary)}
\label{fig:latdetections}
\end{figure}

\begin{table*}[t!]
\begin{center}
\caption{\fermilat\ detection rate for the \twocms\ and the \moj-I Samples (based on a preliminary LAT 11-month catalog, see text)}
\begin{tabular}{|l|c|c|c|c|c|}
\hline \textbf{Set} & \textbf{Total} & \textbf{LAT} & \textbf{LAT Frac.} & \textbf{LAT Frac.} & \textbf{LAT Frac.} \\
& & \textbf{detected} & \textbf{QSO} & \textbf{BL\,Lac} & \textbf{Radio Gal.} \\
\hline 
\moj-I
            & 135 & 85 & 60\% & 86\% & 38\% \\
\hline
\twocms\ & 171 & 82 & 46\% & 79\% & 21\% \\
\hline
\moj-I and \twocms\ & 96 & 63 & 65\% & 84\% & 38\% \\
\hline
\moj-I not \twocms\ & 38 & 22 & 52\% & 100\% & -- \\
\hline
\twocms\ not \moj-I & 79 & 19 & 22\% & 67\% & 8\% \\
\hline
\end{tabular}
\label{table:lat-detrate}
\end{center}
\end{table*}

\section{Results}


Figure \ref{fig:latdetections} shows 
a chart counting the members of the \twocms\ and the \moj-I samples, divided
by optical class, with fill colour for the sources detected 
at the preliminary 11-month \fermilat\ list and transparent colour for the
non-detected one.
Those are shown numerically in Table~\ref{table:lat-detrate}.



\begin{figure}
\includegraphics[width=65mm]{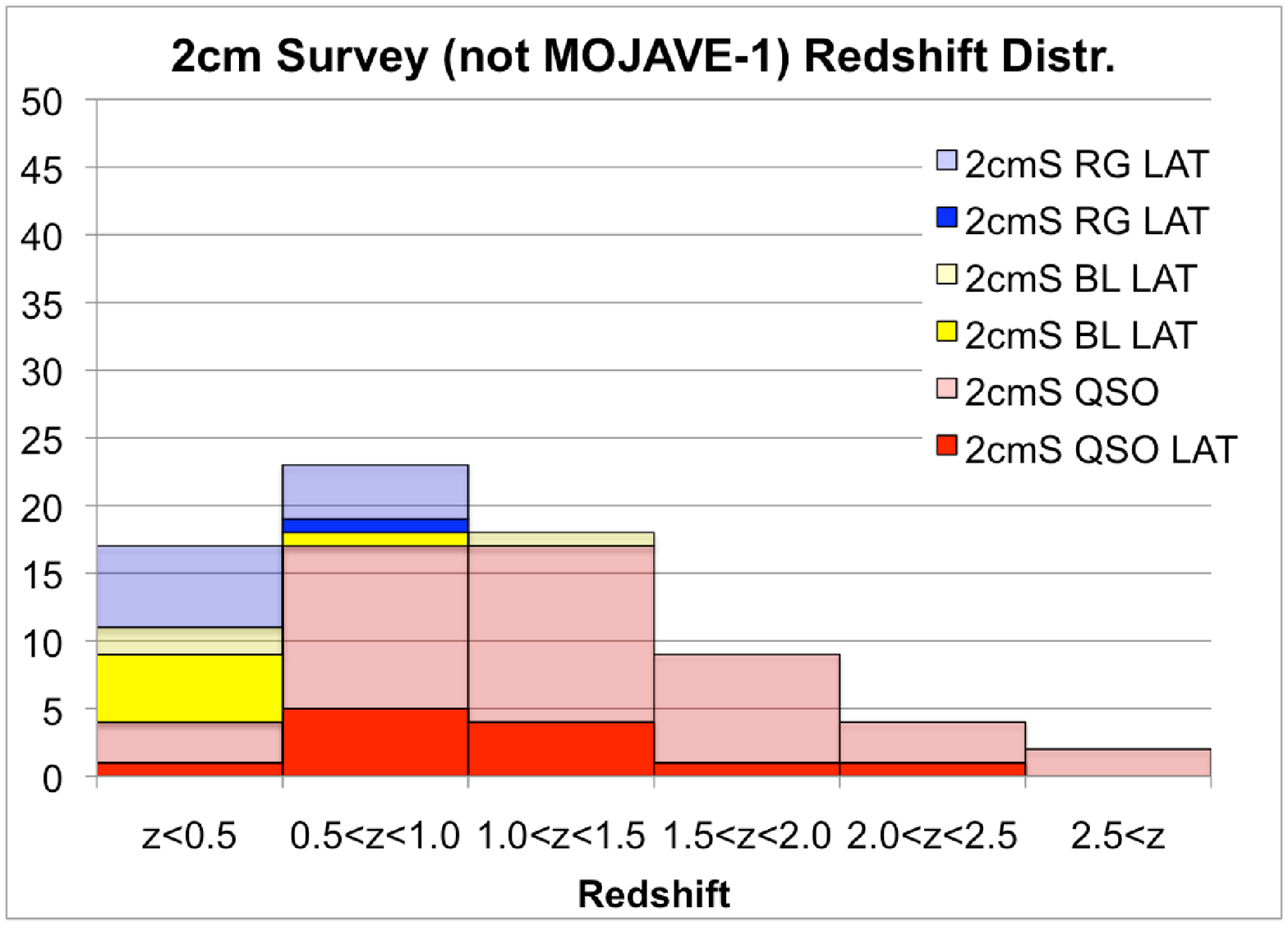}
\includegraphics[width=65mm]{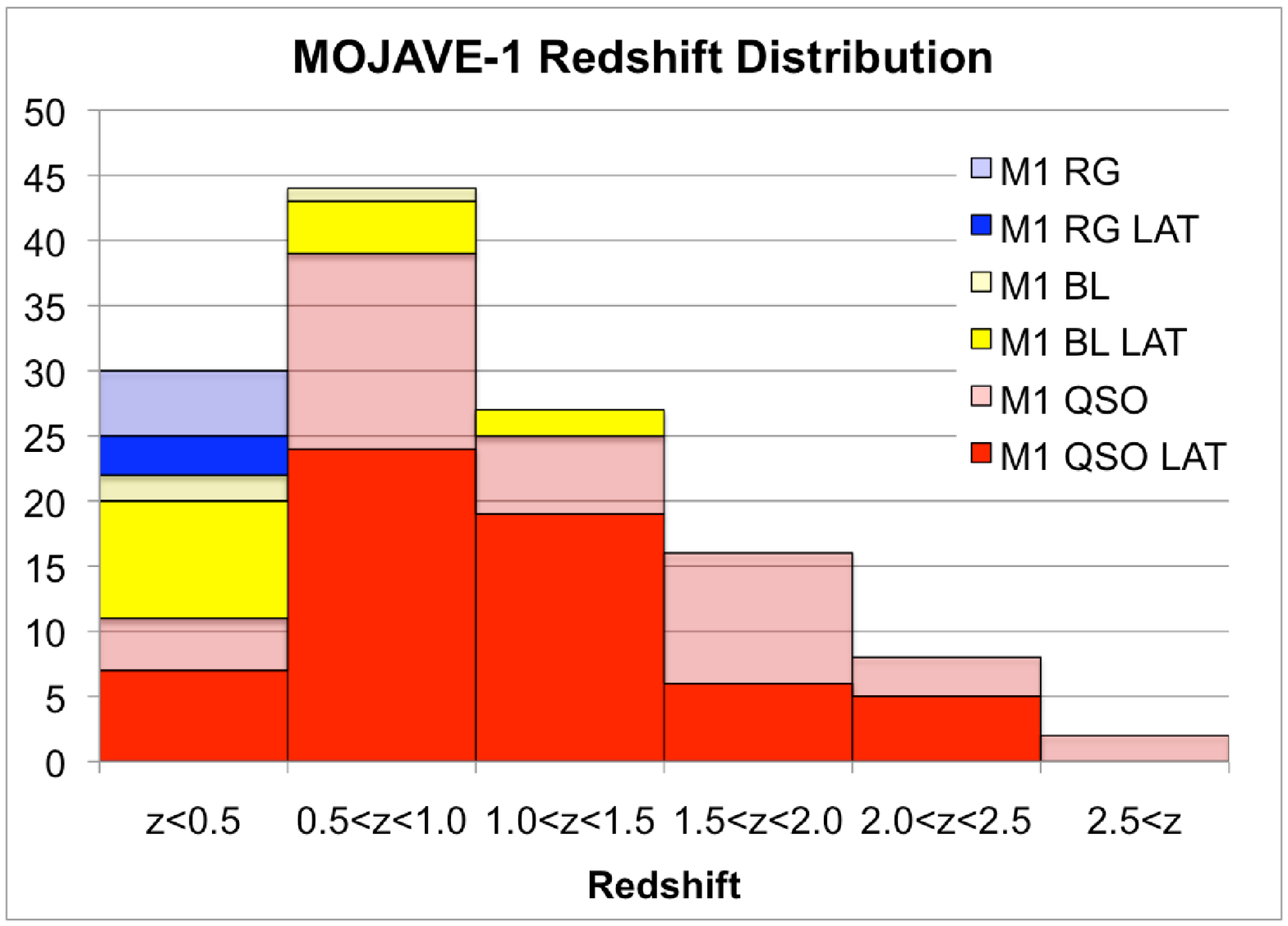}
\caption{Redshift distribution and \fermilat\ detection rate
for the \moj-I (top) and the \twocms\ (not included in \moj-I) 
samples.  Transparent colours represent non-detections, whereas
full colour correspond to LAT detections.  Notice that the used
list of gamma-detections is preliminary.} 
\label{fig:z-dist}
\end{figure}

Figure~\ref{fig:z-dist} shows the distribution of redshifts
and the differences between the statistically complete sample (top)
and the sources from the  \twocms\ not belonging to this (bottom).  
With a smaller number of sources, the fraction of moderate redshift
QSOs is larger for the sources not included in \moj-I.  Again, the 
detection rate is very low for radio galaxies and quasars.

\begin{figure}
\includegraphics[width=65mm]{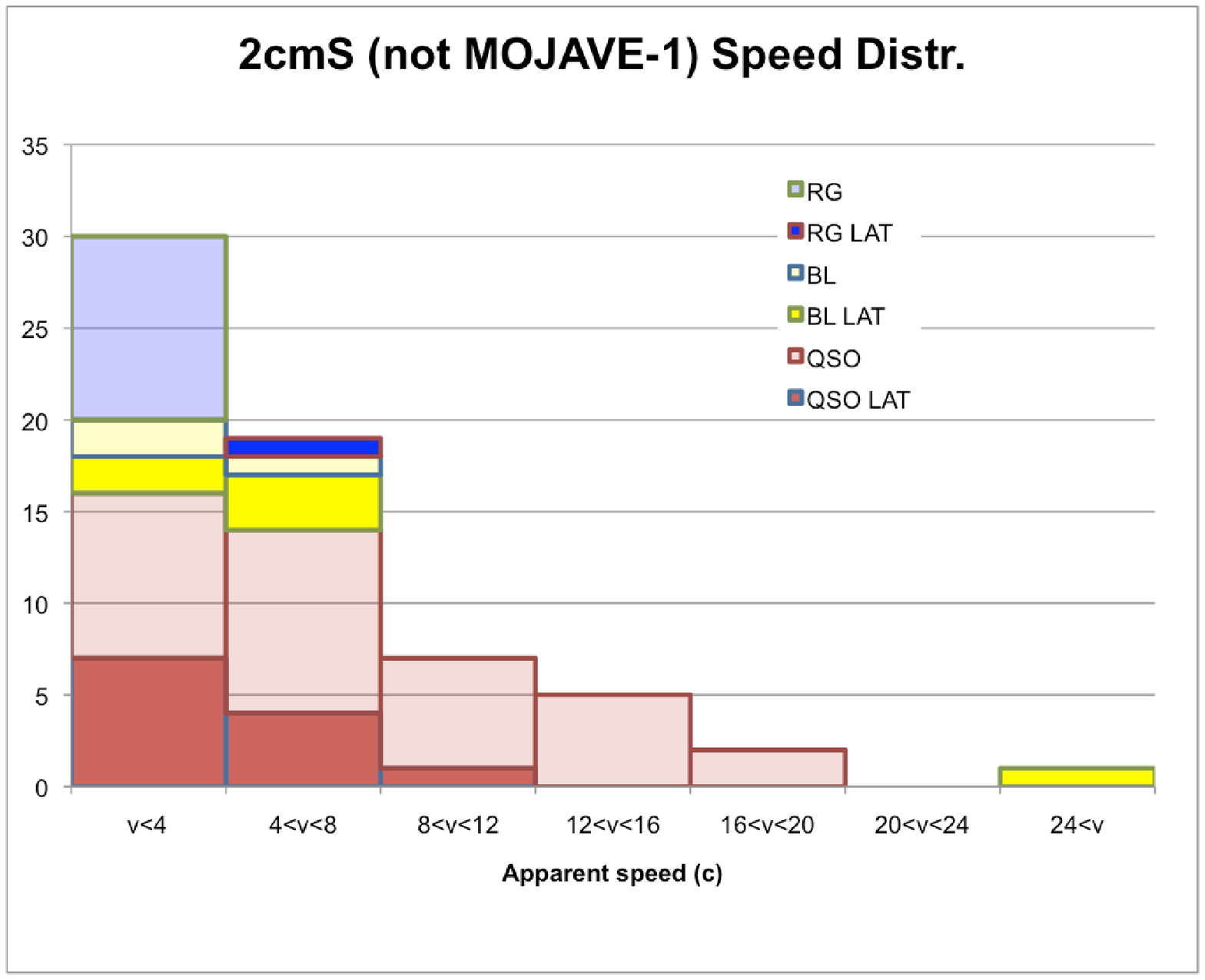}
\includegraphics[width=65mm]{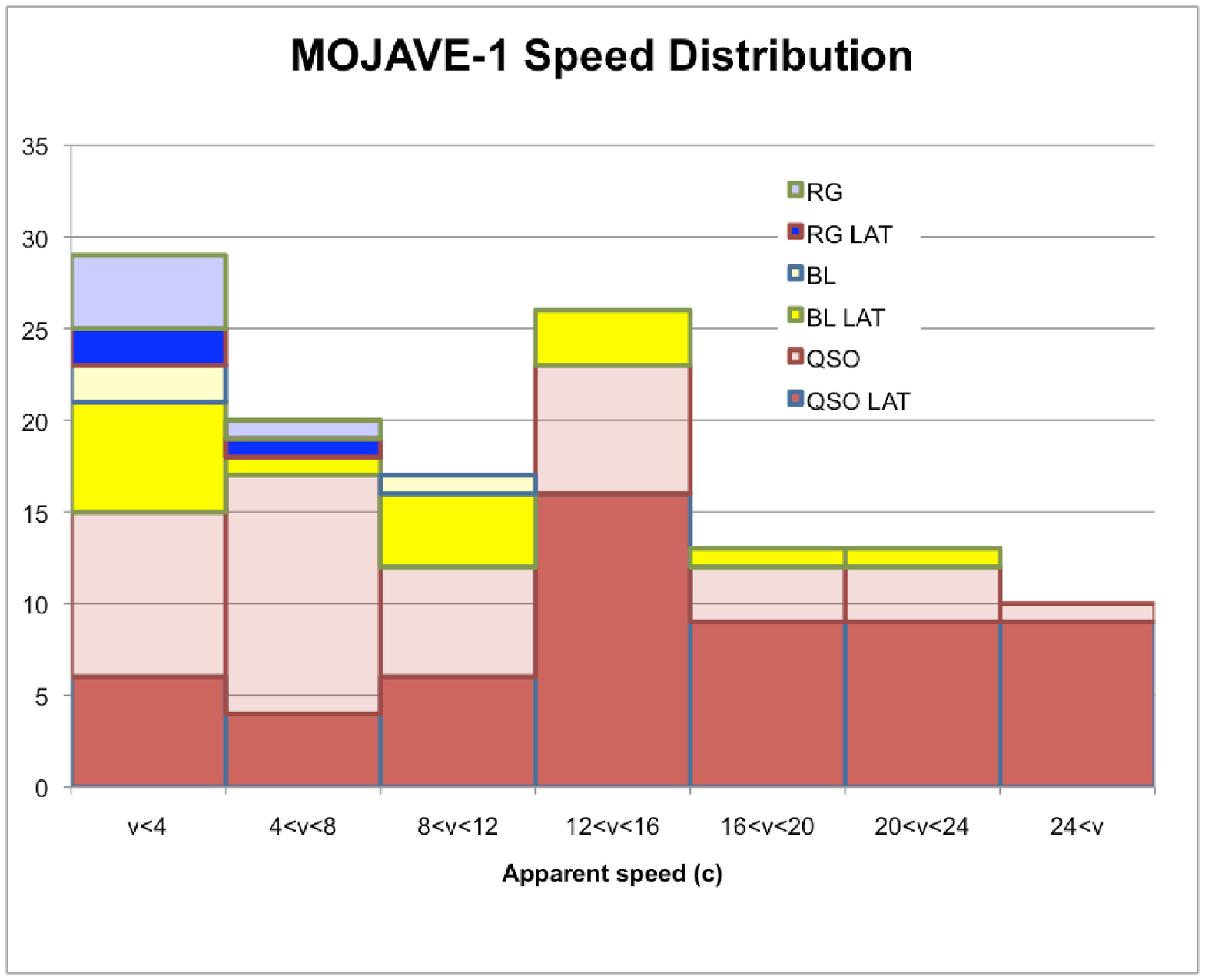}
\caption{Apparent speed distribution and \fermilat\ detection rate for the
\moj-I (top) and the \twocms\ (not included in \moj-I) samples.
LAT-detected sources are shown in full colour and non-detected ones
in transparent colour.  
Notice that the used
list of gamma-detections is preliminary.}
\label{fig:betaapp-dist}
\end{figure}

Figure~\ref{fig:betaapp-dist} shows the distribution of the maximum jet 
speeds for both sets of sources. Notice that the faster jets (above $10c$) 
belong favourably to sources detected by LAT, and that in general all 
sources from the \twocms\ not contained in \moj\ have slower jets.  
From those, the high speed ones are not detected by LAT.  As it
was preliminarly presented by \cite{lis09b} and will be shown
by \cite{pus10}, sources with faster jets tend to be favourably 
detected by LAT.

\section{Discussion and Conclusions}

The gamma detection rate of the \moj\ sample is much 
higher than for the sources of the \twocms\ not belonging to 
the complete sample.  The latter sources have in general slower jets.  
Notice that the \moj-I sample is selected on the base of compact, 
beamed (VLBI) emission, and that the gamma-ray emission is 
correlated with compactness \cite{kov09}.

BL\,Lacs (seen with beamed jets) are more favourably 
detected than QSOs by 
\fermilat, in the same rate for both samples.

The speeds for gamma-detected sources at the \moj-1 sample are 
higher than for the non-detected ones, especially in the case of 
the QSOs.  Notice that the faster the jets are, the more sources 
have been gamma-detected (9 out of 10 for $v>24\,c$).

A 70\% of the quasars of the 2cm Survey not belonging to 
\moj-I were not detected in gamma-rays, which shows a 
big difference in the parent population from the statistically 
complete sample and the additional sources. Notice as well that 
the \moj-1 sample was selected from active sources since the mid 
1990s.  Sources which were active before and not at present would emit 
in gamma-rays less likely.

The kinematic results from the sources belonging to the \twocms\ and
not belonging to the statistically complete sample, and a 
discussion on the \fermilat\ detections from the
newer gamma-catalog to be released in a near future will be presented in 
\cite{ros10}.  In a mid-term, the excellent results to be provided 
by \fermi, including gamma luminosities
and varibility will be tested in the near future together with the radio 
properties, yielding unprecedented information about the nature
of the AGN phenomenon and the emission processes involved.

\bigskip 
\begin{acknowledgments}
In the framework of the MOJAVE collaboration we thank especially
Christian M.\ Fromm for support in the kinematic analysis and 
Chin-Shin Chang for analysis from the \fermilat\ detection list.
We thank the \fermilat\ collaboration for providing preliminary 
gamma-catalog data (11-month list) for the production of this contribution.
The Very Long Baseline Array is operated by the USA National 
Radio Astronomy Observatory, which is a facility of the 
National Science Foundation operated under cooperative agreement by 
Associated Universities, Inc.  
The MOJAVE project is 
supported under USA National Science Foundation grant 0406923-AST.
\end{acknowledgments}

\bigskip 

\end{document}